\documentclass[12pt]{article}
\usepackage{amsfonts,amsmath,amssymb}
\usepackage{color}
\usepackage{epic}
\usepackage{graphics}

\newcommand{\qmbox}[1]{{\qquad\mbox{#1}\quad}}
\def\lddots{\mathinner{\mkern1mu\raise1pt\hbox{.}\mkern2mu
\raise4pt\hbox{.}\mkern2mu\raise7pt\vbox{\kern7pt\hbox{.}}\mkern1mu}}
\makeatletter
\def\numberbysection{\@addtoreset{equation}{section}
\def\theequation{\thesection.\arabic{equation}}}
\makeatother

\numberbysection

\newtheorem{proposition}{Proposition}

\newcommand{\finproof}{{\hfill \rule{5pt}{5pt}}}
\newcommand{\be}{\begin{eqnarray}}
\newcommand{\ee}{\end{eqnarray}}
\newcommand{\non}{\nonumber}

\newcommand{\tr}{\mathop{\rm tr}\nolimits}

\newcommand{\LL}{\mathbb{L}}
\newcommand{\MM}{\mathbb{M}}
\newcommand{\KK}{\mathbb{K}}
\newcommand{\TT}{\mathbb{T}}
\newcommand{\PP}{\mathbb{P}}
\newcommand{\BB}{\mathbb{B}}

\begin{document}

\begin{titlepage}
\vskip 0.4cm
\strut\hfill
\vskip 0.8cm
\begin{center}

{\bf {\Large (Quantum) twisted Yangians:\\symmetry, Baxterisation and centralizers}}\\

\vspace{10mm}

{\large Nicolas Cramp{\'e}\footnote{
crampe@sissa.it}$^{a,b}$ and
Anastasia Doikou\footnote{doikou@bo.infn.it}$^{c}$}

\vspace{10mm}

{\small \emph{$^a$ International School for Advanced Studies,\\
Via Beirut 2-4, 34014 Trieste, Italy}}

{\small \emph{$^b$ Istituto Nazionale di Fisica Nucleare\\
Sezione di Trieste}}

{\small \emph{$^c$ University of Bologna, Physics Department, INFN Section,  \\
Via Irnerio 46, 40126 Bologna, Italy}}

\end{center}

\vfill

\begin{abstract}

Based on the (quantum) twisted Yangians, integrable systems with
special boundary conditions, called soliton non-preserving (SNP),
may be constructed. In the present article we focus on the study
of subalgebras of the (quantum) twisted Yangians, and we show that
such a subalgebra provides an exact symmetry of the rational
transfer matrix. We discuss how the spectrum of a generic transfer
matrix may be obtained by focusing only on two types of special
boundaries. It is also shown that the subalgebras, emerging from
the asymptotics of tensor product representations of the (quantum)
twisted Yangian, turn out to be dual to the (quantum) Brauer
algebra. To deal with general boundaries in the trigonometric case
we propose a new algebra, which also provides the appropriate
framework for the Baxterisation procedure in the SNP case.

\end{abstract}

\vfill
\baselineskip=16pt

\end{titlepage}

\section{Introduction}

It is well established by now that, for any $gl_n$ algebra
or the corresponding quantum deformation (${\cal U}_q(gl_n)$), one may
consider integrable lattice models or quantum field theories with two distinct types of boundary
conditions known as soliton preserving (SP) with underlying
algebra the reflection algebra \cite{cherednik,sklyanin} or soliton non-preserving (SNP)
with underlying algebra the twisted Yangian \cite{twist} or quantum twisted Yangian \cite{moras}.
Historically,
soliton non-preserving boundary conditions were first introduced
in the context of affine Toda field theories on the half line
\cite{cor}, whereas solutions of the (quantum) twisted Yangian first found in \cite{gand}.
Nonetheless up to date, both SP and SNP boundary
conditions have been extensively studied in the context of
integrable quantum spin chains (see e.g. \cite{doikou,our2,doi2}).

In this paper, we focus on the SNP case and we provide a generic
description of the underlying symmetry algebra. The rational and trigonometric cases are considered separately. 
For the
rational case, we investigate the symmetry and we show that
the algebra, emerging from the twisted
Yangian as $|\lambda| \to \infty$, is isomorphic to $so(p,q)$ or $sp(n)$ and
is an exact symmetry of the algebraic open transfer matrix for a particular relation between
the left and right boundaries. In addition, for this particular relation between the boundaries,
we show that it is sufficient to study the spectrum of the transfer matrix for only
two special cases.
Finally, we recall the duality between the Brauer algebra
and this algebra \cite{brauer}.

Similar considerations are made
for the trigonometric case. In this case, the corresponding finite twisted
quantum algebra, obtained from the quantum twisted
Yangian for $|\lambda| \to \infty$, does not seem to provide an exact
symmetry of the corresponding open model. We recall however the
link between the finite twisted quantum algebra and the quantum
Brauer algebra established in \cite{molev} for a trivial boundary.
In the case of a generic boundary, we need to introduce a new algebra.
We prove that a subalebra of this new algebra is the centralizer of the
finite quantum twisted Yangian. The new framework allows us to
find the spectral depending solution via a Baxterisation procedure.

The outline of this paper is as follows: In the next section, we
give the algebraic setting by introducing the FRT presentation \cite{FRT}
for quantum groups and some subalgebras. We show for the rational case that
the set of non local charges emerging from the asymptotic
expansion of the tensor solution of the twisted Yangian, turn
out to provide an exact symmetry of the transfer matrix. We then prove that the spectrum of any
transfer matrix with generic boundary conditions can be deduced from the transfer matrix with special diagonal boundaries. We show that the entailed symmetry algebra commutes with appropriate
representations of the Brauer algebra, depending on the choice of
boundary conditions.
In the last section, the trigonometric case is considered. Surprisingly the emerging boundary quantum algebra is
not an exact symmetry of the trigonometric transfer matrix contrary to the
rational case. Nevertheless, for trivial boundary conditions, we recall
the duality between the quantum Brauer algebra and the twisted
boundary quantum algebra. We finally provide a general framework, which
allows us to deal with generic boundary conditions, and also discuss about the
Baxterisation procedure.

\section{(Quantum) twisted Yangian: general setting}

Let $R(\lambda)$ be a solution of the Yang-Baxter equation
\cite{baxter,zamo,korepin} \be R_{12}(\lambda_{1}-\lambda_{2})\
R_{13}(\lambda_{1})\ R_{23}(\lambda_{2}) =R_{23}(\lambda_{2})\
R_{13}(\lambda_{1})\ R_{12}(\lambda_{1}-\lambda_{2}),\label{YBE} \ee
acting on $\mbox{End}({\mathbb C}^{n})^{\otimes 3}$, and as usual
$~R_{12}(\lambda) = R(\lambda)\otimes {\mathbb I}, 
~~R_{23}(\lambda) = {\mathbb I} \otimes R(\lambda)~$
and so on. In addition, the $R$ matrices we shall deal with
satisfy\\
(i) the unitarity condition\be R_{12}(\lambda)\ R_{21}(-\lambda) \propto
{\mathbb I}  \label{uni} \ee
(ii) the crossing relation \be R_{12}^{t_{1}}(\lambda)\ M_{1}\
R_{12}^{t_{2}}(-\lambda -2i\rho)\ M_{1}^{-1} \propto {\mathbb I},
\label{mm} \ee where $t_{i}$ denotes the transposition on the
$i^{th}$ space\\
(iii) the symmetry \be \label{sym}\Big [M_{1}\ M_{2},\
R_{12}(\lambda) \Big ]=0. \ee Let us also define the matrix
\begin{equation}
\bar
R_{12}(\lambda) \propto R_{12}^{t_{1}}(-\lambda -i\rho)\, ,
\end{equation}
which will
be useful for our purposes here. The proportional factors have no
influence on the definition of the algebra and will be chosen
conveniently.

For an associative infinite algebra $\cal Y$ generated by
$\{L_{ij}^{(m)}|1\leqslant i,j\leqslant n,m=0,1,\dots \}$, one may
define the object \be{\LL}(\lambda)=\sum_{i,j=1}^n E_{ij}\otimes {L}_{ij}(\lambda) \in \mbox{End}({\mathbb C}^{n})
\otimes {\cal Y}[\lambda^{-1}]\, ,\ee where the second  `space' is
occupied by formal series $\displaystyle {L}_{ij}(\lambda)=\sum_{m=0}^{+\infty}
\frac{L_{ij}^{(m)}}{\lambda^m}$ and $E_{ij}$ is a $n$ by $n$ matrix such that
$(E_{ij})_{kl}=\delta_{ik}\delta_{jl}$. The exchange relations may be
written using the so-called FRT relation \cite{FRT}, \be
R_{12}(\lambda_{1} -\lambda_{2})\ {\LL}_{13}(\lambda_{1})\
{\LL}_{23} (\lambda_{2})= {\LL}_{23}(\lambda_{2})\
{\LL}_{13}(\lambda_{1})\ R_{12}(\lambda_{1} -\lambda_{2}),
\label{FRT}\ee where the 1 and 2 stand for the two copies
of $\mbox{End}({\mathbb C}^{n})$ whereas 3, for the space ${\cal
Y}[\lambda^{-1}]$.

In this article, we will be interested in two particular choices of the
algebra called the $gl(n)$ (quantum) Yangian \cite{drinf1, drinf} depending
on the choice of a rational or trigonometric $R$
matrix.

In this type of algebra defined by FRT relation, we can exhibit
different homomorphisms. In particular, we will use
${\LL}(\lambda)\mapsto
\bar{\LL}(\lambda)=f(\lambda){\LL}^t(-\lambda-i\rho)$ i.e. in terms
of generators (entries of ${\LL}$)
\begin{equation}
L_{ij}(\lambda)\mapsto {\bar L}_{ij}(\lambda)=f(\lambda)
L_{ji}(-\lambda-i\rho)
\end{equation}
where $f(\lambda)$ is any function chosen later conveniently.

The (quantum) Yangian defined by relation (\ref{FRT}) are also Hopf algebras.
In particular, they are equipped with a coproduct $\Delta: {\cal Y} \to {\cal Y}
\otimes {\cal Y}$ given by, for $i,~j\in\{1, \ldots, n\}$,
\be \Delta({L}_{ij}(\lambda)) = \sum_{a=1}^n
{L}_{aj}(\lambda) \otimes {L}_{ia}(\lambda)\,
\ee
which can be written also as
\be (\mbox{id} \otimes \Delta){\LL}(\lambda) =
{\LL}_{02}(\lambda)\ {\LL}_{01}(\lambda)\, .\label{co1} \ee
The index 0 stands for the
space $\mbox{End}({\mathbb C}^{n})$, whereas 1 and 2 stand for the
two copies of the algebra. This coproduct induces also
\be
(\mbox{id} \otimes \Delta){\bar\LL}(\lambda) =
{\bar\LL}_{01}(\lambda)\ {\bar\LL}_{02}(\lambda)\, .\label{cob1}
\ee
The $\ell$-coproduct
$\Delta^{(\ell)}: {\cal Y} \to {\cal Y}^{\otimes \ell}$ is defined by the following iteration
$\Delta^{(\ell)} = \left(\mbox{id} \otimes \Delta^{(\ell-1)}\right)
\Delta$ with $\Delta^{(2)}=\Delta$.

In order to deal with integrable systems with non trivial
boundary conditions, one has to consider appropriate subalgebras of $\cal
Y$. For this purpose, we define \be {\mathbb K}_0(\lambda)=\sum_{i,j=1}^n E_{ij}\otimes {K}_{ij}(\lambda)=
{\LL}_{01}(\lambda)\ {\cal K}_0(\lambda) \bar {\LL}_{01}(\lambda),
\label{gensol} \ee with ${\cal K}(\lambda)$ a c-number matrix
solution of the following equation
\begin{equation} R_{12}(\lambda_{1}-\lambda_{2})\ {{\cal K}}_{1}(\lambda_{1})\
\bar R_{12}(\lambda_{1}+\lambda_{2})\ {{\cal K}}_{2}(\lambda_{2}) =
{{\cal K}}_{2}(\lambda_{2})\ \bar R_{12}(\lambda_{1}+\lambda_{2})\
{{\cal K}}_{1}(\lambda_{1})\ R_{12}(\lambda_{1}-\lambda_{2})
\label{re}
\end{equation}
The generators encompassed in ${\mathbb K}(\lambda)$ generate a
subalgebra $\cal T$ of $\cal Y$, whose exchange relations are
given by (\ref{re}), where now ${\cal K}(\lambda)$ is replaced
by ${\mathbb K}(\lambda)$. Depending on the choice of the $R$ matrix,
$\cal T$ is the (quantum)
twisted Yangian\footnote{In the case of the twisted Yangian, a
supplementary relation is required for $\KK$ (see (\ref{symt})) }
\cite{twist,moras}. It is clear that $\KK(\lambda)$ allows the
expansion in powers of $\lambda^{-1}$ as we shall see in subsequent
section, providing explicit forms of the generators of the twisted
Yangian.

The subalgebra ${\cal T}$ is not a Hopf algebra but has a
structure of co-ideal inherited essentially from the (quantum)
Yangian, $\Delta: {\cal T} \to {\cal T} \otimes {\cal Y}$, such that
(see also \cite{dema, dema2}), for $i,~j\in\{1, \ldots, n\}$, \be \Delta({K}_{ij}(\lambda)) =
\sum_{a,b=1}^n{K}_{ab}(\lambda) \otimes {L}_{ia}(\lambda) \bar {
L}_{bj}(\lambda) \label{coc} \ee
or, equivalently, $(\mbox{id} \otimes \Delta){\KK}(\lambda)=
{\LL}_{02}(\lambda){\LL}_{01}(\lambda)\ {\cal K}_0(\lambda)
\bar {\LL}_{01}(\lambda)\bar {\LL}_{02}(\lambda)$.
One can
exploit the existence of tensor product realizations of the
(quantum) twisted Yangian in order to build the corresponding
quantum system that is the open quantum spin chain with SNP boundary
conditions. We define 
\be \TT_{0} (\lambda) &=& (\mbox{id} \otimes
\Delta^{(N)}) {\LL}(\lambda) = {\LL}_{0N}(\lambda)\ldots {\LL}_{01}
(\lambda)\, ,\\
\bar \TT_{0} (\lambda) &=& (\mbox{id} \otimes
\Delta^{(N)}) \bar{\LL}(\lambda) = \bar{\LL}_{01}(\lambda)\ldots
\bar{\LL}_{0N} (\lambda)\, . \label{tt'} \ee
Then, the general tensor type solution of the (\ref{re}) takes the
form
\be {\BB}_{0}
(\lambda) =(\mbox{id} \otimes
\Delta^{(N)}) {\KK}(\lambda)= \TT_{0}(\lambda)\ {\cal K}^{(R)}_{0}(\lambda)\ \bar
\TT_{0}(\lambda)\, , \label{skl} \ee
where ${\cal
K}^{(R)}_{0}(\lambda)$ is a c-number matrix solution of (\ref{re})
interpreted as the reflection on the right boundary. The entries
$B_{ij}(\lambda)$ of the matrix $\BB(\lambda)$ can be computed
explicitly by $ {B}_{ij}(\lambda) =\Delta^{(N)}({K}_{ij}(\lambda))$.

Finally, we introduce the transfer matrix of the open spin chain
\cite{sklyanin}, which may be written as
\be t(\lambda) = \tr_{0}\
\Big \{ {\cal K}_{0}^{(L)}(\lambda)\ {\BB}_{0}(\lambda) \Big \}\, ,
\label{transfer} \ee
where ${\cal K}^{(L)}(\lambda)={\cal
K}(-\lambda-i\rho)$ encodes the interaction with the left boundary and ${\cal
K}(\lambda)$ is a solution of (\ref{re}). It can be shown
\cite{sklyanin, doikou}, using the fact that ${\BB}(\lambda)$ is a solution
of (\ref{re}), that transfer matrix (\ref{transfer}) provides
family of commuting operators i.e., \be \Big [t(\lambda),\
t(\lambda')\Big ] =0\, . \label{com} \ee The latter commutation
relation (\ref{com}) ensures the integrability of the relevant
models.

\section{The rational case}

One of the main objectives of the present study is the derivation of the exact
symmetry for integrable models associated to (quantum) twisted
Yangian. We shall first examine the rational case and, in the
subsequent section, we shall proceed with the trigonometric case.

The $gl_{n}$ $R$ matrix is written in the following simple
form,\be R(\lambda) = {\mathbb I} +{i \over \lambda}{\cal P}
\label{r0}\ee where  $\displaystyle {\cal P} = \sum_{a,b=1}^n
E_{ab}\otimes E_{ba}$ is the permutation
operator, acting on $({\mathbb C}^{n})^{\otimes 2}$. Note that
for the rational case $\rho = {n \over
2}$ and $M={\mathbb I}$ (see (\ref{mm})). Let us define
$\check{\cal  P} =\rho{\mathbb I} -{\cal Q} $ where
$\displaystyle {\cal Q} = \sum_{a,b=1}^n E_{ab}\otimes E_{ab}$
is a one-dimensional projector satisfying \be {\cal Q}\ {\cal P} = {\cal P}\
{\cal Q}= {\cal Q}, ~~~~~{\cal Q}^{2} = n {\cal Q}, \ee and
consequently $\check {\cal P}^{2} = \rho^{2} {\mathbb I}$. 
Then, the $\bar R$ matrix can be written as
\be\bar R(\lambda)  ={\mathbb I} + { i \over \lambda}
\check {\cal P} \label{rb}\ee
For the rational case, a solution of the fundamental equation
(\ref{FRT}) is provided by evaluation map i.e. \be {L}_{ij}(\lambda)
= \delta_{ij} + {ie_{ji} \over \lambda}, ~~~~~\bar
{L}_{ij}(\lambda)= \frac{\lambda+i\rho}{\lambda}{L}_{ji}(-\lambda
-i\rho) = \delta_{ij} +{i\rho\delta_{ij}-ie_{ij} \over \lambda}
\label{l}\ee where $e_{ij}$ are the generators of the Lie algebra
$gl(n)$ satisfying
\begin{equation}
[{e}_{ij},{e}_{kl}]=\delta_{jk}{e}_{il}-\delta_{il}{e}_{kj}
\end{equation}
Let us introduce ${\mathbb P}=\sum_{i,j} E_{ij}\otimes E_{ji}$ and
$\check {\mathbb P} = \sum_{i,j} E_{ij}\otimes (\rho\delta_{ij}-
E_{ij})$. Then, we can write
\begin{equation}
\LL(\lambda)=1+\frac{i\PP}{\lambda}\qmbox{and}{\bar\LL}(\lambda)=1+\frac{i{\check\PP}}{\lambda}
\end{equation}
It is clear that ${\cal P}$ and $\check {\cal P}$ are the fundamental
representations of the corresponding ${\mathbb P}$ and $\check {\mathbb
P}$.

\subsection{Symmetry of the transfer matrix}

We shall also need in what follows the $c$--number solution of
(\ref{re}) for the rational case, which is any constant matrix
($\lambda$ independent) such that ${\cal K} =\pm {\cal K}^t$
\cite{our2, MRS}. We shall show that different choices of $\cal K$
provide different symmetry algebras. We suppose also that ${\cal K}$ is
invertible and with real entries. We recall that there does not
exist invertible antisymmetric $n\times n$ matrix for n odd
(Jacobi's theorem easily checked by $\det({\cal K})=(-1)^n\det({\cal
K})$). For this choice of $R$ matrix, the elements encompassed in
$\BB(\lambda)$ generated the twisted Yangian \cite{twist}. The
elements satisfied the following supplementary symmetry
relation\footnote{The differences between the relation presented
here and this one in \cite{twist} are due to the shift in the
spectral parameter of the matrix $\bar R(\lambda)$ in the
commutation relation (\ref{re}) and to the factor in $\bar
\LL(\lambda)$. }
\begin{equation}
\label{symt}
\BB^t(\lambda)=\pm\BB(-\lambda-i\rho)g(\lambda)+\frac{i}{2\lambda+i\rho}
\big(\BB(-\lambda-i\rho)g(\lambda)-\BB(\lambda)\big),
\end{equation}
where the upper (resp. lower) sign corresponds to ${\cal K}$
symmetric (resp. antisymmetric) and
$g(\lambda)=\left(\frac{\lambda+i\rho}{\lambda}\right)^{2N}$.

We shall consider here the expansion of ${\BB}(\lambda)$ as $\lambda
\to \infty$ in order to obtain explicit expressions of generators of
the twisted Yangian (see also \cite{doikouy}). We shall keep up to ${1\over \lambda}$ terms
which is sufficient
for our purposes. Bearing in mind the expansions of ${\LL}(\lambda)$ and $\bar
{\LL}(\lambda)$, we may easily deduce the asymptotic behavior of ${\BB}(\lambda)$ as
$\lambda \to \infty$, i.e.\be {\BB}_0(\lambda \to \infty) = {\cal
K}_0+{i\over \lambda}{\BB}_0^{(1)} + \ldots ={\cal K}_0+{i\over
\lambda}\sum_{i=1}^{N}({\cal K}_0\  \check {\mathbb P}_{0i} +{\mathbb
P}_{0i}\ {\cal K}_0) +\ldots\, . \label{asyt2} \ee Let us define the
following combination emerging essentially from the asymptotic
expansion of the generalized solution of (\ref{re}) \be {\mathbb
K}^{(1)}=  {\cal K}\ \check {\mathbb P} + {\mathbb P}\ {\cal
K}\qmbox{i.e} {K}^{(1)}_{ab}=\rho{\cal
K}_{ab}-\sum_{c=1}^n\left({\cal K}_{ac}e_{cb}-e_{ca}{\cal
K}_{cb}\right) . \label{k0}\ee The non--local charges i.e. the
entries of (\ref{asyt2}), may be written simply as coproducts of the
twisted Yangian elements namely ${\BB}^{(1)} = \Delta^{(N)}({\mathbb
K}^{(1)})$. In particular, the asymptotic expansion of (\ref{coc})
provides
\begin{eqnarray}
\Delta({\mathbb K}^{(1)})&=&{\mathbb K}^{(1)}\otimes {\mathbb I} +
{\mathbb I} \otimes {\KK}^{(1)}\ =\ {\cal K}_0\ \check {\mathbb
P}_{01} + {\mathbb P}_{01}\ {\cal K}_0+ {\cal K}_0\check {\mathbb
P}_{02} + {\mathbb P}_{02}{\cal K}_0\;.
\end{eqnarray}
The elements $K_{ab}^{(1)}$ are called primitive and, in particular,
their coproduct is cocomutative.

In the following, we will study more precisely the algebra ${\cal
T}^{(1)}$ spanned by $B_{ab}^{(1)}$ but, before that, we shall show
that they commute with the transfer matrix of the system provided
that special boundary conditions are considered. In the proof we
shall simply exploit the underlying algebraic relations provided by
(\ref{re}) as $\lambda_{1} \to \infty$ i.e.
\begin{equation}
\big[\ \BB_1^{(1)}\ ,\ \BB_2(\lambda)\ \big]={\cal
P}_{12}\big(\BB_1(\lambda){\cal K}_2-{\cal
K}_1\BB_2(\lambda)\big)+{\cal K}_1{\cal
Q}_{12}\BB_2(\lambda)-\BB_2(\lambda){\cal Q}_{12}{\cal K}_1\, .
\end{equation}
Then, we extract the element in the position $(a,b)$ in the space 1
and $(c,d)$ in the space 2 \be \big[\ {B}_{ab}^{(1)}\ ,\
{B}_{cd}(\lambda)\ \big]={\cal K}_{ad}\ {B}_{cb}(\lambda)-{\cal
K}_{cb}\ {B}_{ad}(\lambda)+{\cal K}_{ac}\ {B}_{bd}(\lambda) -{\cal
K}_{db}\ {B}_{ca}(\lambda)\ . \label{alg} \ee From the
latter exchange relations it is entailed that for a generic ${\cal
K}^{(R)}$ matrix the elements ${B}^{(1)}_{ab}$ do not commute with
the transfer matrix. If however we consider the case where the left
boundary is ${\cal K}^{(L)} = {{\cal K}^{(R)}}^{-1}$, the transfer
matrix becomes  \be t(\lambda) = \sum_{l,\ c=1}^{n}{\cal
K}^{(L)}_{lc}\ {B}_{cl}(\lambda)\, . \ee From the latter equation
and bearing in mind the exchange relations (\ref{alg}), it is easy to
show that \be \Big [t(\lambda),\ {B}^{(1)}_{ab} \Big] =0. \ee
Note that an alternative proof may be formulated along the lines of \cite{doikou, done}.
The
particular choice of the left boundary indicates that the two
boundaries of the chain have to be appropriately tuned so that all
the elements ${B}^{(1)}_{ab}$ commute with the open transfer matrix.
In fact, the special cases: (i) ${\cal K}^{(L)}_{ab}
=\delta_{ab}={\cal K}^{(R)}_{ab}$,
(ii) ${\cal K}^{(L)}_{ab} = \delta_{a,n+1-b}={\cal K}^{(R)}_{ab}$,
(iii) ${\cal K}^{(L)}_{ab} = (-1)^a\ i\ \delta_{a,n+1-b}=
{\cal K}^{(R)}_{ab}$ fall to the category above. The symmetry of
cases (ii) and (iii) has been studied
also in \cite{doi2}. One may extract valuable information concerning
the exact symmetry of a chain with generic boundary conditions
exploiting the algebraic relations (\ref{alg}).

Now, we describe more precisely the algebra ${\cal T}^{(1)}$
spanned by $\{B_{ab}^{(1)}|1\leq a,b \leq n\}$. We are
going to show that in the case $\cal K$ symmetric it is isomorphic
to the Lie algebra $so(p,q)$ whereas for $\cal K$ antisymmetric it
is isomorphic to $sp(n)$. For convenience, we introduce the sign
$\epsilon=\pm$ such that ${\cal K}^t=\epsilon {\cal K}$. By
expanding expressions (\ref{symt}) and (\ref{alg}) up to
$\lambda^{-1}$, we get
\begin{eqnarray}
\label{lie1}
&&B_{ba}^{(1)}=-\epsilon B_{ab}^{(1)}+\epsilon 2\rho N {\cal K}_{ab}\\
\label{lie2} &&\big[\ {B}_{ab}^{(1)}\ ,\ {B}_{cd}^{(1)}\ \big]={\cal
K}_{ad}\ {B}_{cb}^{(1)}-{\cal K}_{cb}\ {B}_{ad}^{(1)}+{\cal K}_{ac}\
{B}_{bd}^{(1)} -{\cal K}_{db}\ {B}_{ca}^{(1)}\ .
\end{eqnarray}
Let us now introduce the following $n \times n$ matrices
\begin{equation}
\label{ge} {\cal
G}^+=\mbox{diag}(\underbrace{1,\dots,1}_p,\underbrace{-1,\dots,-1}_q)\qmbox{and}
{\cal G}^-=\mbox{diag}(\underbrace{1,\dots,1}_{n/2})\otimes
\left(\begin{array}{cc}0& 1\\-1&0\end{array}\right)\ ,
\end{equation}
where $p+q=n$ and the second case is valid only for $n$ even. A
well-known linear algebra result is that the matrices ${\cal
G}^\epsilon$ are the normal forms of the matrix $\cal K$ over reals
under congruence i.e there exists an invertible real matrix $\cal U$
such that
\begin{equation}
{\cal U}{\cal K}{\cal U}^t={\cal G}^\epsilon \ .
\end{equation}
The mapping
\begin{eqnarray}
\BB^{(1)}&\longmapsto&\epsilon  {\cal U} (\BB^{(1)}-\rho N {\cal
K}){\cal U}^t{\cal G}^\epsilon={\cal U}\BB^{(1)}{\cal K}^{-1}{\cal
U}^{-1}-\rho N=\MM
\\
\mbox{i.e.}~~B_{ab}^{(1)}&\longmapsto&
\epsilon\sum_{\alpha,\beta,\gamma=1}^n{\cal U}_{a
\alpha}\big(B_{\alpha\beta}^{(1)}-\rho N{\cal
K}_{\alpha\beta}\big){\cal U}_{\gamma \beta}{\cal
G}^\epsilon_{\gamma b}=M_{ab}
\end{eqnarray}
defines an algebra isomorphism from ${\cal T}^{(1)}$ to  $so(p,q)$
(resp. to $sp(n)$) for $\epsilon=+$ (resp. $\epsilon=-$). The
bijection is proven by the fact that $\cal U$ is invertible. By
direct computation of the relations satisfying by $M_{ab}$ starting
from relations (\ref{lie1}) and (\ref{lie2}), we recognize the
following defining relations of $so(p,q)$ (resp. $sp(n)$)
\begin{eqnarray}
\MM^t\ {\cal G}^\epsilon&=&-{\cal G}^\epsilon\ \MM\  \\
{}[\MM_1,\MM_2]&=&[\MM_2,{\cal P}_{12}\ - \ {\cal G}_1^\epsilon\
{\cal Q}_{12}\ ({\cal G}_1^\epsilon)^{-1}]
\end{eqnarray}
which show the algebra homomorphism.

\subsection{Treatment of general boundary condition}

Henceforth we focus on the fundamental
representation i.e. $\LL_{0i}(\lambda)\mapsto R_{0i}(\lambda)$ and
${\bar\LL}_{0i}(\lambda)\mapsto {\bar R}_{0i}(\lambda)$ and we
restrict ourselves to the case where ${\cal K}^{(R)}={{\cal
K}^{(L)}}^{-1}=\cal K$. Then, we get the following representation for ${\mathbb B}(\lambda)$,
\be
{\mathbb B}_0(\lambda)\mapsto R_{0N}(\lambda)\dots R_{01}(\lambda)\, {\cal K}_0\,
\bar{R}_{01}(\lambda)\dots \bar{R}_{0N}(\lambda)={\cal B}_0(\lambda)\, .\ee

In this case, we have the following important proposition which
allows us to find the spectrum of the transfer matrix for any
boundary condition by studying only the case where the boundary
conditions are given by ${\cal G}^\pm$.
\begin{proposition}
Let $\cal K$ be any solution of (\ref{re}) and $t_{\cal
K}(\lambda)=\tr_0({\cal K}^{-1}_0 {\cal B}_0(\lambda))$. There exists invertible matrix
$\cal U$ such that ${\cal U} {\cal K} {\cal U}^t={\cal G}^\epsilon$
where ${\cal G}^\epsilon$ are defined in (\ref{ge}). Let $$t_{{\cal
G}^\epsilon}(\lambda)= \tr_0\left(({\cal G}^\epsilon_0)^{-1} R_{0N}(\lambda)\dots
R_{01}(\lambda)\, {\cal G}^\epsilon_0\, \bar{R}_{01}(\lambda)\dots \bar{R}_{0N}(\lambda)\right)\, .$$\\ Then
$t_{\cal K}(\lambda)$ and $t_{{\cal G}^\epsilon}(\lambda)$ have the
same eigenvalues, their eigenvectors (say $V_{\cal K}$ and $V_{{\cal
G}^\epsilon}$ respectively) being related trough
\begin{equation}
V_{{\cal G}^\epsilon}= {\cal U}_1\dots {\cal U}_N V_{\cal
K}\ .
\end{equation}
\end{proposition}
\textit{Proof.} The proof of this proposition is based on the relation
\begin{equation}
 {\cal U}_1\dots {\cal U}_N\ t_{\cal K}(\lambda)=t_{{\cal
G}^\epsilon}(\lambda)\ {\cal U}_1\dots {\cal U}_N
\end{equation}
which is obtained using the property $[\bar{R}_{0i}(\lambda),({\cal
U}_0^t)^{-1}{\cal U}_i]=0$ and the symmetry relation (\ref{sym}). \finproof\\

This result is similar to the one obtained in the soliton preserving
case in \cite{our2,lepetit,martin,our0}. This proposition allows us to restrict the study of the spectrum of
the transfer matrix only to the cases where the boundaries are given
by ${\cal G}^\epsilon$. Let us emphasize that although the
eigenvalues are identical the models obtained from $t_{\cal K}$ and
$t_{{\cal G}^\epsilon}$ may be different.

\subsection{Link with the Brauer algebra \label{brauer}}

We shall now focus on the relation between the symmetry algebra and
the Brauer algebra \cite{brauer}. A well-known presentation of the classical
Brauer algebra ${\cal B}_{N}(\delta)$ is realized by $2N-2$
generators $\sigma_{i}$ and $\tau_{i}$ ($1\leq i \leq N-1$) obeying
exchange relations:
\be
\sigma_{i}^{2} &=& 1, ~~~~~\tau_{i}^{2} =
\delta\ \tau_{i}, ~~~~~\sigma_{i}\ \tau_{i} =
\tau_{i}\ \sigma_{i}= \tau_{i}, ~~~~~i=1, \ldots, N-1 \non\\
\sigma_i\ \sigma_j &=& \sigma_j\ \sigma_i, ~~~~~\tau_i\ \tau_j =
\tau_j\ \tau_i, ~~~~~\sigma_i\ \tau_j = \tau_j\ \sigma_i,
~~~~~~|i-j|>1 \non\\ \sigma_i\ \sigma_{i+1}\ \sigma_{i} &=&
\sigma_{i+1}\ \sigma_i\ \sigma_{i+1}, ~~~~~\tau_{i}\ \tau_{i\pm 1}\
\tau_{i} =\tau_{i}, \non\\ \sigma_{i} \tau_{i+1}\ \tau_{i}
&=&\sigma_{i+1}\ \tau_{i}, ~~~~~\tau_{i+1}\ \tau_{i}\ \sigma_{i+1} =
\tau_{i+1}\ \sigma_{i}, ~~~~~i=1, \ldots N-2 \label{bra}
\ee
By direct computation, it is clear that the following map
\be\label{repbrauer}
\sigma_{i}
\mapsto \epsilon\ {\cal P}_{i\ i+1} \qmbox{and} \tau_{i} \mapsto
{\cal K}_i\ {\cal Q}_{i\ i+1}\ {\cal K}_i^{-1} \ ,
\ee
is a representation of the Brauer algebra ${\cal B}_{N}(n)$ for any matrix ${\cal K}$
satisfying ${\cal K}^t=\epsilon \cal K$ with $\epsilon=\pm$.

By representing relation (\ref{asyt2}), the conserved quantities in the fundamental
representation are given by
\be
{\cal B}^{(1)}_0=\sum_{j=1}^N\Big( {\cal P}_{0j}\
{\cal K}_0+{\cal K}_0\ \check{\cal P}_{0j}\Big)
\ee
and generate an algebra
isomorphic to $so(p,q)$ and $sp(n)$ depending on the
choice of the $\cal K$ matrix. Using relation (\ref{repbrauer}), it is now straightforward to show
that the conserved quantities are the centralizers of the Brauer
algebra i.e.
\be
[\epsilon\ {\cal P}_{i\ i+1}\ ,\ {\cal B}^{(1)}_0]=0 \qmbox{and} [
{\cal K}_i\ {\cal Q}_{i\ i+1}\ {\cal K}_i^{-1}\ ,\  {\cal B}^{(1)}_0]=0\, .
\ee

A few comments are in order at this point. In the SP case, the
presence of a non trivial right boundary modifies naturally the form
of the non-local charges, nevertheless they still commute with the
transfer matrix as long as the left boundary is trivial.
Furthermore, these charges continue to be the centralizers of an
extended Hecke algebra called the B-type Hecke algebra (see
\cite{doikou3}). In the case we examine here, the
boundary non-local charges consist a symmetry algebra for the
transfer matrix as long as the left and right boundaries are closely
interrelated and are also centralizers of the Brauer algebra. In
this spirit, it seems pointless to consider a `boundary' extension of
the Brauer algebra analogously to the Hecke case. However it is
possible to conceive a generalization of the classical Brauer
algebra regarding the ${\cal K}$ matrix as a representation of the
extra element of the `extended' algebra. Indeed, let us consider a
possible extension of the Brauer algebra by introducing two
supplementary generators ${\mathrm b}$ and ${\mathrm b}^{-1}$,
satisfying
\begin{eqnarray} &&{\mathrm b}\ {\mathrm b}^{-1}={\mathrm
b}^{-1}\ {\mathrm b} = 1\ ,~~~~~\sigma_j\ b=b\ \sigma_j,
~~~~~\tau_j\ b=b\ \tau_j\qmbox{for}j\geq 2 \\
&&\sigma_{1}\ {\mathrm b}\ \sigma_{1}\ {\mathrm b} = {\mathrm b}\
\sigma_{1}\ {\mathrm b}\ \sigma_{1} ,\\
&&\sigma_{1}\ {\mathrm b}\ \tau_{1}\ {\mathrm b} = {\mathrm b}\
\tau_{1}\ {\mathrm b}\ \sigma_{1}, ~~~~~\sigma_{1}\ {\mathrm
b}^{-1}\ \tau_{1}\ {\mathrm b}^{-1} = {\mathrm b}^{-1}\ \tau_{1}\
{\mathrm b}^{-1}\ \sigma_{1}\ .
\end{eqnarray}
It is easy to prove that the following map \be\sigma_{i} \mapsto
\epsilon {\cal P}_{i\ i+1} \qmbox{,} \tau_{i} \mapsto {\cal Q}_{i\
i+1}\  \qmbox{,} {\mathrm b} \mapsto {\cal K}_1 \qmbox{and}
{\mathrm b}^{-1} \mapsto {\cal K}_1^{-1}\ee is a representation of
this extended Brauer algebra. In the SP case, the representation
of the extended Hecke algebra is important since it provides by
Baxterisation new solutions of the reflection algebra. In our
case, unfortunately, the Baxterisation does not apply given that
no numerical solution depending on the spectral paremeter of (\ref{re}) exists.

The picture presented above offers a rather tempting interpretation
of the $\cal K$ matrix  in terms of the `boundary' generator. However, another
interpretation exists where the $\cal K$ matrix allows us to define new
interaction in the bulk. Let us define, for any ${\cal K}^t=\epsilon
\cal K$, \be \bar R_{12}'(\lambda) &=& {\cal K}_{1} \bar
R_{12}(\lambda) {\cal K}^{-1}_{1}= {\cal K}_{2}\bar R_{12}(\lambda)
{\cal K}^{-1}_{2}\, .  \ee The transfer
matrix, in the case where $({\cal K}^{(L)})^{-1} ={\cal K}^{(R)} =
{\cal K}$, may be written as
\be t_{\cal K}(\lambda) =\tr_0 \{ R_{0N}(\lambda)\ldots R_{01}(\lambda)  \bar
R'_{01}(\lambda) \ldots \bar R'_{0N}(\lambda)
 \}\, . \ee
 The
commutativity of the transfer matrix (which provides, as usual, the integrability of the system) is ensured by
\begin{equation} R_{12}(\lambda_{1}-\lambda_{2})\ {\cal B}'_{1}(\lambda_{1})\
\bar R'_{12}(\lambda_{1}+\lambda_{2})\ {\cal B}'_{2}(\lambda_{2})=
{\cal B}'_{2}(\lambda_{2})\ \bar R'_{12}(\lambda_{1}+\lambda_{2})\
{\cal B}'_{1}(\lambda_{1})\ R_{12}(\lambda_{1}-\lambda_{2}) \label{re2}
\end{equation} where ${\cal B}'_0(\lambda)={\cal B}_0(\lambda)\, {\cal K}_0^{-1}$. The crucial
point here is that the first non trivial terms in the expansion of ${\cal B}'_0(\lambda)$  for
$\lambda\rightarrow+\infty$ i.e.
\be
\label{gsp}
{\cal B}'^{(1)}_0=\sum_{j=1}^N\Big( {\cal P}_{0j}+
{\cal K}_0\ \check{\cal P}_{0j} {\cal K}_0^{-1}\Big)=N\rho+\sum_{j=1}^N\Big( {\cal P}_{0j}-
{\cal K}_0\ {\cal Q}_{0j}\  {\cal K}_0^{-1}\Big)
\ee
still provides the symmetry of the transfer matrix. In addition, it is still
the centralizer of the representation of the Brauer algebra, i.e.
\be
[\epsilon\ {\cal P}_{i\ i+1}\ ,\ {\cal B'}^{(1)}_0]=0 \qmbox{and} [
{\cal K}_i\ {\cal Q}_{i\ i+1}\ {\cal K}_i^{-1}\ ,\  {\cal B'}^{(1)}_0]=0\,
~~~~i\in \{1, \ldots, N-1 \}. \ee
This construction is more symmetrical in the sense that the same operator
appears in the symmetry algebra (\ref{gsp}) and in the representation of
the Brauer algebra (\ref{repbrauer}).

Within this frame of mind, it seems rather unnecessary to discuss about a
`boundary extension' of the Brauer algebra. To conclude
we showed the duality between the symmetry algebra of the transfer
matrix and the Brauer algebra for suitable representations,
depending on the choice of boundaries.

\section{The trigonometric case}

The ${\cal U}_{q}(\widehat{gl}_{n})$ $R$ matrix derived in
 \cite{jimbo} may be written in a compact form as \be R(\lambda) =
2\sinh  (\lambda+i\mu)\ {\cal P} + 2\sinh ( \lambda)\  {\cal P}\ U
~~~\mbox{and} ~~~~U = \sum_{\underset{i\neq j}{i,j =1}}^{n}(E_{ij} \otimes
E_{ji} - q^{-sgn(i-j)} E_{ii} \otimes  E_{jj})\ ,
\label{r} \ee
where $q=e^{i\mu}$.
In this case, $\rho=\mu n/2$ and the matrix $M$
(\ref{mm}) is defined  as \be  &&  M_{ij}=q^{n-2j+1}\
\delta_{ij}\, .\label{M} \ee
Defining $R_{12}[q]=R_{12}={\cal P}(q+U)$, we can write the $R$ matrix more symmetrically as follows
\be
\label{rsym}
R(\lambda)= e^\lambda R_{12}-e^{-\lambda}R_{21}^{-1}=e^\lambda R_{12}[q]-e^{-\lambda}R_{12}[q^{-1}]^{t_1t_2}
\ee
The $\bar R$ matrix is given by
\be
\bar R(\lambda) =e^{-\lambda-i\rho} R_{12}^{t_1}-e^{\lambda+i\rho}(R_{21}^{-1})^{t_1}
\ee
In particular, we get $\bar R(-i\rho) = (q-q^{-1}){\cal Q}$.
It is well-known that rational $R$ matrix is a limit of trigonometric $R$ matrix. Indeed, rescaling the spectral
parameter $\lambda\rightarrow \mu \lambda$ in definition (\ref{r}) of the trigonometric $R$ matrix and taking the
limit $\mu\rightarrow 0$, we get rational $R$ matrix (\ref{r0}). This limit, called the scaling limit, will be useful
to compare the results of sections 3 and 4.

A solution of equation (\ref{FRT}), where now $R$ is a
trigonometric matrix given above, may be written in the following simple
form (see also \cite{ft11}) \be {\LL}(\lambda)=
e^{\lambda} {\LL}^{+} - e^{-\lambda} {\LL}^{-}, ~~~~~
{\LL}^{+} = \sum_{\underset{i\leq j}{i,j=1}}^{n} E_{ij} \otimes \ell^+_{ij},
~~~~~{\LL}^{-} = \sum_{\underset{i\geq j}{i,j=1}}^{n} E_{ij} \otimes
\ell_{ij}^{-}. \label{lh0} \ee with the matrices ${\LL}^{+}$
(${\LL}^{-}$) apparently upper (lower) triangular and
$\ell^+_{ij}$, $\ell_{ij}^{-}$ the generators of the finite quantum group $U_{q}(gl_{n})$.
Its exchange relations can be also written using the FRT presentation as follows
\be
R_{12}\LL^\pm_1\LL^\pm_2&=&\LL^\pm_2\LL^\pm_1 R_{12}\\
R_{12}\LL^+_1\LL^-_2&=&\LL^-_2\LL^+_1 R_{12}.
\ee
We have the following additional constraints on diagonal elements, for $1\leq i \leq n$,
\be\ell^+_{ii}\ell^-_{ii}=\ell^-_{ii}\ell^+_{ii}=1\, .\label{unil}\ee
Similarly to the rational case, we introduce
$\bar {\LL}(\lambda) = {\LL}(-\lambda -i\rho)^{t}$ which can be written as
\be \bar {\LL}(\lambda)= e^{\lambda} \bar {\LL}^{+} - e^{-\lambda} \bar
{\LL}^{-}, ~~~~~\bar {\LL}^{+} = \sum_{\underset{i\leq j}{i,j=1}}^{n}
E_{ij} \otimes \bar \ell^+_{ij}, ~~~~~{\LL}^{-} = \sum_{\underset{i\geq j}{i,j=1}}^{n}  E_{ij} \otimes \bar \ell_{ij}^{-}\, , \label{lh1}\ee where
$\bar \ell^\pm_{ij}=-e^{\pm i\rho}\ \ell^\mp_{ji}$.

In \cite{twisty}, a classification of numerical solutions of equation (\ref{re})
for trigonometric $R$ matrix has been presented. More precisely, the general invertible
solutions $\cal K(\lambda)$, up to a global factor, can be written as
\begin{equation}
\label{DD}
{\cal K}(\lambda)={\cal D} {\cal G}(\lambda) {\cal D}
\end{equation}
where ${\cal D}$ is any invertible constant matrix and $\cal G(\lambda)$ is one of the following matrices
\begin{enumerate}
\item[(i)] ${\cal G}(\lambda)=\mathbb{I}$
\item[(ii)]
$\displaystyle\frac{ {\cal G}(\lambda)}{\left(q^{1\over 2}+q^{-{1\over 2}}\right)}=e^{\lambda} q^{n \over 4}\left({q^{-1} \over q^{-1}+1}
\sum_{i=1}^n \epsilon_i^2 E_{ii} +
\sum_{\underset{i<j}{i,j}=1}^n \epsilon_i\epsilon_jE_{ij}\right)\pm e^{-\lambda}q^{-{n \over 4}}\left({q \over q+1}
\sum_{i=1}^n \epsilon_i^2 E_{ii} +
\sum_{\underset{i>j}{i,j=1}}^n \epsilon_i\epsilon_jE_{ij}\right)$
where $\epsilon_i=\sqrt{(-1)^{in}}$ and by convention $\sqrt{1}=1,\ \sqrt{-1}=i$.
The solution given here is linked to the solution given in \cite{twisty} by transformation
(\ref{DD}) where $\displaystyle {\cal D}=\sum_{i=1}^n \sqrt{(-1)^{in}}E_{ii}$ and by a global factor.
\item[(iii)] $\displaystyle{\cal G}(\lambda) = e^{2\lambda}q^{n-1 \over 2}E_{1n} -e^{-2\lambda}q^{-{n-1 \over 2}}
E_{n1} + \sum_{i=1}^{{n-2\over 2}}(q^{1\over 2}E_{2i\ 2i+1} -q^{-{1\over 2}}E_{2i+1\ 2i}), ~~~~n ~~~\mbox{even}$.
\item[(iv)]
$\displaystyle{\cal G}(\lambda) = \sum_{i=1}^{{n\over 2}}(q^{1\over 2}E_{2i-1\ 2i}
-q^{-{1\over 2}}E_{2i\ 2i-1})={\cal G}_q^-, ~~~~n ~~~\mbox{even}$.
\end{enumerate}
Notice that the solutions (i), (ii) coincide with previously known
results found in \cite{gand,dema}. By the scaling limit,
solutions (i) and (iv) becomes respectively $\cal G^+$ (for $q=0$)
and $\cal G^-$ (see relation (\ref{ge})). Solution (iii) becomes
$E_{1n}-E_{n1}+\sum_{i=1}^{\frac{n-2}{2}}E_{2i,2i+1}-E_{2i+1,2i}$
which is antisymmetric. Solution (ii), with the upper sign,
becomes $\sum_{i,j=1}^n E_{ij}$ which is never invertible and,
with the lower sign, $\sum_{i<j}(E_{ij}-E_{ji})$ which is
antisymmetric. We recover different invertible solutions of the
reflection equation with the rational $R$ matrix and some non
invertible solutions.

Finally, let us remark that each solution can be written as
\be
\label{Gsym}
{\cal G}(\lambda)={\cal G}^+(\lambda,q)+\, \sigma\,{\cal G}^+(-\lambda,q^{-1})^t
\ee
where ${\cal G}^+(\lambda,q)$ is equal to (i) $\mathbb{I}/2$ ($\sigma=+$),
(ii) $e^{\lambda} q^{n \over 4}\left({q \over q+1}
\sum_{i=1}^n\epsilon_i^2 E_{ii} + \sum_{i<j}\epsilon_i\epsilon_j E_{ij}\right)$ ($\sigma=\pm$),
(iii) $e^{2\lambda}q^{n-1 \over 2}E_{1n}+ \sum_{i=1}^{{n-2\over 2}}q^{1\over 2}E_{2i\ 2i+1}$ ($\sigma=-$)
and (iv) $\sum_{i=1}^{{n\over 2}}q^{1\over 2}E_{2i-1\ 2i}$ ($\sigma=-$).
Note the similarity between relations (\ref{rsym}) and
(\ref{Gsym}) and the fact that ${\cal G}^+(\lambda,q)$ is upper diagonal as $R_{12}$.

\subsection{Finite subalgebras \label{secf}}

The asymptotic expansion of ${\mathbb B}(\lambda)$ as $\lambda \to \pm \infty$
will provide explicit expressions of the finite quantum twisted Yangian
generators. We shall keep $\lambda$ independent expressions,
bearing in mind the form of ${\LL}$ (\ref{lh0}) and $\bar {\LL}$ (\ref{lh1}) and also the generic solutions given above.
It is clear that the first three solutions preserve the triangular
decomposition of the ${\mathbb B}(\lambda \to \pm\infty) =
{\mathbb B}^{\pm}=\sum_{i,j=1}^nE_{ij}\otimes B^\pm_{ij}$ matrix, while for the last one ${\mathbb B}^{\pm}$ is
not triangular any more. Explicit expressions of ${\mathbb B}^+$ for
the various solutions (i)--(iv) are given below
\begin{itemize}
\item[(i)]
For $i< j$, $
B_{ij}^+ =
\displaystyle\sum_{k=i}^j \ell_{ik}^{+}\ \bar \ell^+_{kj}$ , $
B_{ii}^+ =-e^{i\rho}$  and for $i>j$, $
B_{ij}^+ = 0$.
\item[(ii)]
For $i<j$, $B_{ij}^+ \propto\displaystyle\frac{q^{-1}}{q^{-1}+1}
\sum_{k=i}^j\epsilon^2_k \ell_{ik}^{+}\ \bar \ell^+_{kj}+
\sum_{k=i}^{j-1}\sum_{p=k+1}^j\epsilon_k\epsilon_p\ell_{ik}^{+}\
\bar \ell^+_{pj}$,
$\displaystyle B_{ii}^+ \propto -\frac{\epsilon_i^2 e^{i\rho}\, q^{-1}}{q^{-1}+1}$  and  for $i>j$,
$B_{ij}^+ = 0$
\item[(iii)]$B_{1n}^+\propto \ell_{11}^+\bar\ell^+_{nn}$ and $0$ otherwise.
\item[(iv)]
$\displaystyle B^{+}_{ij} \propto q^{1 \over 2}\ \sum_{\frac{i+1}{2}\leq k
\leq \frac{j}{2}}  \ell^{+}_{i,2k-1}\ \bar \ell^+_{2k,j} - q^{-{1 \over 2}} \sum_{\frac{i}{2}\leq k
\leq \frac{j+1}{2}} \ell_{i,2k}^+\ \bar \ell_{2k-1,j}^+$ for $i \leq j+1$ and $0$ otherwise.
\end{itemize}
We have used property (\ref{unil}) and the relations between $\ell$ and $\bar \ell$.
Similarly, it is possible to compute explicitly the charges $B_{ij}^-$ in terms
of $\ell^-_{ij}$ and $\bar \ell^-_{ij}$.

The charges $B_{ij}^{\pm}$ form an algebra, called ${\cal T}_f$ ($f$ for finite), with
the following defining exchange relations emerging from (\ref{re})
as $\lambda_{i} \to \pm \infty$, i.e. we get
\be
\label{fi1}R_{12}\ {\mathbb B}_{1}^{+}\ \bar R_{12}\ {\mathbb B}_{2}^{+} &=&
{\mathbb B}_{2}^{+}\ \bar R_{12}\ {\mathbb B}_{1}^{+}\ R_{12}\, \\
\label{fi2}R_{12}\ {\mathbb B}_{1}^{-}\ R_{12}^{t_1}\ {\mathbb B}_{2}^{-} &=&
{\mathbb B}_{2}^{-}\ R_{12}^{t_1}\ {\mathbb B}_{1}^{-}\ R_{12}\, \\
\label{fi3}R_{12}\ {\mathbb B}_{1}^{+}\ \bar R_{12}\ {\mathbb B}_{2}^{-} &=&
{\mathbb B}_{2}^{-}\ \bar R_{12}\ {\mathbb B}_{1}^{+}\ R_{12}\, \\
\label{fi4}R_{12}\ {\mathbb B}_{1}^{+}\ R_{12}^{t_1}\ {\mathbb B}_{2}^{-} &=&
{\mathbb B}_{2}^{-}\ R_{12}^{t_1}\ {\mathbb B}_{1}^{+}\ R_{12}\,
\ee
where $\bar R_{12}=(R_{21}^{-1})^{t_1}$. In \cite{doikou3, doikoun}, analogous results have been obtained in the SP case.

\textbf{Remarks}
(1) The subalgebra, denoted ${\cal T}_f^+$,
generated by $B_{ij}^+$ is in fact isomorphic to the one, denoted ${\cal T}_f^-$,
generated by $B_{ij}^-$:
it is proved using the invertible transformation $\check{\mathbb B}_1^+=M_{1}
\left((\mathbb B_{1}^+)^{-1}\right)^t$ and showing that $\check{\mathbb B}^+$ satisfies
relation (\ref{fi2}).

(2) When $\cal K(\lambda)$ is the solution (i) (resp. solution (iv)), the subalgebra
${\cal T}^-_f$ is the twisted quantum algebra $U'_q(so_n)$
(resp. $U'_q(sp_n)$): these algebras were introduced in \cite{gav,noumi}.

(3) For the solution (iii), there are only two non-vanishing generators
$B^+_{1n}$ and $B^-_{n1}$. In this case, relations (\ref{fi1})-(\ref{fi4}) reduce
only to $[B^+_{1n},B^-_{n1}]=0$ which is an abelian algebra and is non-deformed.

(4) The algebra associated to the solution (ii) was discussed in \cite{dema}, but here is treated in detail
and gives interesting non trivial results as we shall see in the subsequent sections.

In the following, we will not study the case (iii) since it provides a trivial finite algebra.

\subsection{Symmetry of the transfer matrix}

We shall present in what
follows the exchange relations among the finite algebras studied previously
and the entries of the transfer matrix
(i.e. $\tr_0 \BB_0(\lambda)$).
Such relations will essentially manifest the existing --if any-- symmetries of
the model. More precisely, let us consider (\ref{re}) for
$\lambda_{1} \to \pm \infty $. Bearing in mind that $R_{12}(\lambda \to
+\infty) \propto R_{12}$ and $\bar R(\lambda \to +\infty) \propto \bar
R_{12}=(R_{21}^{-1})^{t_1}$, one obtains the following set of exchange relations:
\be
R_{12}\BB_1^+ \bar R_{12} \BB_2(\lambda)= \BB_2(\lambda)\bar R_{12}\BB_1^+R_{12}. \label{exc1}\ee
Similar expressions may be derived for the case that $\lambda \to
-\infty$, but they are omitted here for brevity. Explicit
expressions for the ${\cal U}_{q}(gl_{3})$ case are provided in
Appendix \ref{appexp}.

Let us focus on the charges next to diagonal, their explicit
expressions for solution (ii) coincide with the ones found in
\cite{dema} i.e. \be B_{i,i+1}^{+} = \ell_{ii}^{+}\ \bar
\ell_{i,i+1}^{+} + \ell_{i,i+1}^+ \bar \ell_{i+1,i+1}^{+} + (q+1)
\ell_{ii}^+\ \bar \ell_{i+1,i+1}^+. \ee
For ${\cal K} ={\mathbb I}$, the expression is the same with the
last term vanishing. Let $c = q-q^{-1}$. Then, from
(\ref{exc1}), we obtain:
\be c \sum_{l} E_{i+1,l}
\otimes B_{il}(\lambda) -q c \sum_{l}E_{i,l} \otimes B_{i+1,l}(\lambda) + \sum_{l,j}f_{j} E_{jl}\otimes B_{i,i+1}^+
\ B_{jl}(\lambda) \non\\ = c \sum_{l} E_{li}
\otimes B_{l,i+1}(\lambda) -q c \sum_{l} E_{l,i+1}
\otimes B_{l,i}(\lambda)  + \sum_{l,j}\bar f_{j} E_{lj}
\otimes B_{lj}(\lambda)\ B^+_{i,i+1}\, .
\ee
From the latter
equations, one concludes that $B^+_{i,i+1}$ do not commute
with the SNP transfer matrix for any choice of left boundary (see
also Appendix \ref{appexp} for the full set of exchange relations for the
${\cal U}_{q}(gl_{3})$ case) contrary to the SP case and to the
rational SNP case where the exchange relations of the (\ref{exc1})
allow the study of symmetry of the transfer matrix with
boundary conditions. The higher
order expansion of the transfer matrix provides naturally
non-local quantities, which due to integrability commute with the
transfer matrix, however such quantities being in involution form
an abelian algebra. The main challenge within this context is the
search of a non abelian algebra, which at the same time would form a
symmetry of the transfer matrix as in the case of SP boundary
conditions.

\subsection{Quantum Brauer algebra\label{secbrauer}}

In this subsection, we basically recall the results obtained in \cite{molev}.
The quantum deformation of the Brauer algebra ${\cal B}_{N}(z,q)$
\cite{molev} is generated by $\sigma_{1}, \ldots
,\sigma_{N}, \tau_{1}$ subject to the following defining relations
\be  \sigma_{i}^{2}
&=& (q-q^{-1})\sigma_{i} +1, ~~~~~\sigma_{i}\ \sigma_{j}
=\sigma_j\ \sigma_i ,~~~~~\sigma_i\ \sigma_{i+1}\ \sigma_{i} =
\sigma_{i+1}\ \sigma_i\ \sigma_{i+1},  \non\\  \tau_{1}^{2} &=&
{z- z^{-1} \over q -q^{-1}} \tau_{1}, ~~~~~
\sigma_{1}\ \tau_{1}= \tau_{1}\ \sigma_{1} = q \tau_{1} \non\\
\tau_{1}\ \sigma_{2}\ \tau_{1} &=& z\ \tau_{1}, ~~~~~\sigma_{i}\
\tau_{1} = \tau_{1}\ \sigma_{i},  ~~~~~i=2, \ldots N-1 \non\\
\tau_{1}\ (zq\ \zeta^{-1} &+& z^{-1}q^{-1} \zeta)\ \tau_{1}\
(q\zeta^{-1}+q^{-1} \zeta )= (q\ \zeta^{-1}+q^{-1} \zeta )\
\tau_{1}\ (z\ q\ \zeta^{-1} +z^{-1}q^{-1} \zeta)\ \tau_{1}
\label{qbra} \ee
where $\zeta = \sigma_{2}\ \sigma_{3}\ \sigma_{1}\ \sigma_{2}$.
Notice that the generators $\{\sigma_{i}\}$ form the usual Hecke algebra
${\cal H}_{N}(q)$. It has been shown in \cite{molev} that the map
\be \label{repB}\sigma_{i} \mapsto U_{i,i+1}+q={\cal P}_{i,i+1}R_{i,i+1}=\hat R_{i,i+1}\qmbox{and}\tau_{1} \mapsto  {\cal P}^{t_{1}}_{12}\,
M_{1}^{-1}\,  \ee
is a representation of ${\cal B}_{N}(q^n,q)$. The difference between the
algebra defined here and the one introduced in \cite{molev} is such that, although we chose a different definition for the coproduct ${\mathbb T}(\lambda)$, the quantum Brauer algebra is still the centralizer of the quantum twisted Yangian.

As in the rational case, we focus on the fundamental representation of
$U_q(\widehat {gl}_n)$ acting on $({\mathbb C}^n)^{\otimes N}$ i.e.
\be
\TT(\lambda)\mapsto R_{0N}(\lambda)\dots R_{01}(\lambda)\, ,
\ee
which provides a representation of $\BB(\lambda)\mapsto R_{0N}(\lambda)\dots R_{01}(\lambda)\, {\cal K}_0(\lambda)\,
\bar R_{01}(\lambda)\dots \bar R_{0N}(\lambda)$. In particular, we get
\be
\label{repTp}
\BB^+_0&\mapsto& R_{0N}\dots R_{01}\, {\cal K}^+_0\, \bar R_{01}\dots \bar R_{0N}=B^+_0\, \\
\label{repTm}
\BB^-_0&\mapsto& R_{N0}^{-1}\dots R_{10}^{-1}\, {\cal K}^-_0\, R_{01}^{t_0}\dots R_{0N}^{t_0}=B^-_0\,
\ee
where ${\cal K}^\pm={\cal K}(\lambda\rightarrow\pm \infty)$ and we recall that $\bar R_{0i}=(R_{i0}^{-1})^{t_0}$.

A.I. Molev, in \cite{molev}, focused on the solution (i) (${\cal K}^{\pm}={\mathbb I}$) i.e. $U'_q(so_n)$
and proved, in this case, that the actions of the quantum Brauer algebra
${\cal B}_{N}(q^n,q)$ (see (\ref{repB})) on the
space $({\mathbb C}^n)^{\otimes N}$ and the one of the algebra ${\cal T}^-_f$, generated by $B^-_{ij}$, (see
(\ref{repTm})) commute with each other i.e.
\be
[U_{i,i+1}+q\, ,\, B^-_0]=0\qmbox{and}[{\cal P}^{t_{1}}_{12}\, M_{1}^{-1}\, ,\, B^-_0]=0\, .
\ee
We can prove similarly the same result for the whole
algebra ${\cal T}_f$ (generated by $B^\pm_{ij}$).

Unfortunately, it seems impossible to prove similar results for all the solutions ${\cal K}(\lambda)$. In addition, the matrices $\bar R$ and ${\cal K}^\pm$ cannot be written just using matrices similar to the ones representing $\sigma_i$ and $\tau_i$ whereas it has been possible in the rational case (see end of section \ref{brauer}). These remarks indicate that we need a more general framework we will introduce in the following section.

\subsection{Generalized quantum Brauer algebra and Baxterisation}

It is well known that the trigonometric $R$ matrix can be
obtained from generators of the Hecke algebra: such an operation is called Baxterisation. Indeed, starting from $\hat R_{i,i+1}$ satisfying Hecke algebra,
the following linear combination
\be\label{rs}
\hat R_{i,i+1}(\lambda)=e^\lambda \hat R_{i,i+1} - e^{-\lambda} \hat R_{i,i+1}^{-1}
\ee
satisfies the braided Yang-Baxter equation
\be
\label{ybeb}
\hat R_{12}(\lambda_1-\lambda_2)\hat R_{23}(\lambda_1)\hat R_{12}(\lambda_2)
=\hat R_{23}(\lambda_2)\hat R_{12}(\lambda_2)\hat R_{23}(\lambda_1-\lambda_2)\ee and unitarity condition
$\hat R(\lambda)\hat R(-\lambda)\propto {\mathbb I}$.
As explained before, it does not seem
possible to construct $\bar R(\lambda)$ and ${\cal K}(\lambda)$ in a similar fashion. However, there exist equivalent forms where the Baxterisation for these two matrices becomes possible.

Let us introduce the conjugate index $\bar k=n+1-k$,
the matrix $V=q^{k-\frac{N+1}{2}}E_{k\bar k}$ for $n$ odd and
$V=i(-1)^kq^{k-\frac{N+1}{2}}E_{k\bar k}$ for $n$ even (such that $V^2={\mathbb I}$ and $M=V^tV$) and
$\widetilde \BB_0(\lambda)=\BB_0(\lambda)V_0^t$. Then, we get from equation (\ref{re}), the following braided reflection equation
\be
\hat R_{12}(\lambda_1-\lambda_2)\widetilde \BB_1(\lambda_1)\widetilde R_{12}(\lambda_1+\lambda_2)
\widetilde \BB_1(\lambda_2)=\widetilde \BB_1(\lambda_2)\widetilde R_{12}(\lambda_1+\lambda_2)
\widetilde \BB_1(\lambda_1)\hat R_{12}(\lambda_1-\lambda_2)\label{ret}
\ee
where
\be
\widetilde R_{12}(\lambda)={\cal P}_{12}V_2^t\bar R_{12}(\lambda)V_2^t
={\cal P}_{12}V_2^tR_{12}^{t_1}(-\lambda-i\rho)V_2^t
\ee
Obviously, the algebras generated by $\widetilde \BB_0(\lambda)$ and $\BB_0(\lambda)$ are isomorphic.
Whereas for $\bar R(\lambda)$ there is no symmetric form, the matrix $\widetilde R(\lambda)$ can be
written as
\be
\label{trs}
\widetilde R_{12}(\lambda)=e^{-\lambda-i\rho} \widetilde R_{12}-e^{\lambda+i\rho} \widetilde R_{12}^{-1}
\ee
where $\widetilde R_{12}={\cal P}_{12}V_2^t R_{12}^{t_1}V_2^t$.

All numerical solutions $\widetilde{\cal K}(\lambda)$ of equation (\ref{ret}) can be obtained from the classification ${\cal K}(\lambda)$ using the relation $\widetilde{\cal K}(\lambda)={\cal K}(\lambda)V^t$.
The only solution which gives a non trivial finite algebra (see section \ref{secf}) and depends on the spectral parameter is the solution (ii). The solution $\widetilde{\cal K}(\lambda)$ corresponding to the cases (ii) can be also written symmetrically as
\be\label{tks}\widetilde{\cal K}(\lambda)=
e^\lambda q^{\frac{n}{4}} \widetilde{\cal K}\pm e^{-\lambda}q^{-\frac{n}{4}}\widetilde{\cal K}^{-1}\ee
where $\displaystyle \widetilde{\cal K}=\left(q^{1\over 2}+q^{-{1\over 2}}\right)\left({q^{-1} \over q^{-1}+1}
\sum_{i} \epsilon_i^2 E_{ii} +
\sum_{i<j} \epsilon_i\epsilon_jE_{ij}\right)V^t$.

We are now interested in the reciprocity: we want to find an algebra whose representation gives $\hat R$,
$\widetilde R$ and $\widetilde{\cal K}$ and which is sufficient to prove the different relations (Yang-Baxter
equation, unitarity, reflection equation) that the matrices $\hat
R(\lambda)$,
$\widetilde R(\lambda)$ and $\widetilde{\cal K}(\lambda)$ of the forms (\ref{rs}), (\ref{trs}) and (\ref{tks}) must satisfy. Let us define a new algebra, called
${\cal N}_N(w,\pm,q)$, generated by $\{\sigma_i\}$ satisfying the relation of the Hecke algebra
and by the invertible generators $\{\rho_{i}\}$ and $\{b_{i}\}$
subject to the following defining relations,
\be
&&\rho_i(\rho_i-\rho_i^{-1})=\pm w^2(\rho_i-\rho_i^{-1})\label{bd1}
\\
&&\sigma_i b_j=b_j\sigma_i\qmbox{,}\rho_i b_j=b_j\rho_i\qmbox{for $i>j$ or $i<j-1$}\\
&&\sigma_i\rho_j=\rho_j\sigma_i\qmbox{,}\rho_i\rho_j=\rho_j\rho_i\qmbox{for $|i-j|>1$}\\
&&\sigma_{i+1}\rho_i\rho_{i+1}=\rho_i\rho_{i+1}\sigma_i\qmbox{,}
\sigma_{i}\rho_{i+1}\rho_{i}=\rho_{i+1}\rho_{i}\sigma_{i+1}\,\\
&&\sigma_{i+1}\rho_{i}^{-1}\rho_{i+1}=\rho_{i}\rho_{i+1}^{-1}\sigma_{i}\qmbox{,}
\sigma_{i}\rho_{i+1}^{-1}\rho_{i}=\rho_{i+1}\rho_{i}^{-1}\sigma_{i+1}\\
&&
\sigma_{i}\rho_{i+1}\rho_{i}^{-1}-\sigma_{i}^{-1}\rho_{i+1}^{-1}\rho_{i}=\rho_{i+1}^{-1}\rho_{i}\sigma_{i+1}-\rho_{i+1}\rho_{i}^{-1}\sigma_{i+1}^{-1}\\
&&
\sigma_{i+1}\rho_{i}\rho_{i+1}^{-1}-\sigma_{i+1}^{-1}\rho_{i}^{-1}\rho_{i+1}=\rho_{i}^{-1}\rho_{i+1}\sigma_{i}-\rho_{i}\rho_{i+1}^{-1}\sigma_{i}^{-1}\\
&&\sigma_ib_i\rho_i^{-1}b_i=b_i\rho_i^{-1}b_i\sigma_i
\qmbox{,}\rho_i b_i\sigma_ib_i=b_i\sigma_ib_i\rho_i\\
&&\sigma_ib_i\rho_ib_i-\sigma_ib_i^{-1}\rho_i^{-1}b_i^{-1}=b_i\rho_ib_i\sigma_i
-b_i^{-1}\rho_i^{-1}b_i^{-1}\sigma_i\\
&&\sigma_ib_i^{-1}\rho_ib_i-\sigma_i^{-1}b_i\rho_ib_i^{-1}=b_i\rho_ib_i^{-1}\sigma_i
-b_i^{-1}\rho_ib_i\sigma_i^{-1}\\
&&\sigma_ib_i^{-1}\rho_i^{-1}b_i-\sigma_i^{-1}b_i\rho_i^{-1}b_i^{-1}=b_i\rho_i^{-1}b_i^{-1}\sigma_i
-b_i^{-1}\rho_i^{-1}b_i\sigma_i^{-1}\label{ed2}
\ee
We define also two quotients of ${\cal N}_N(w,\pm,q)$, called ${\cal N}^0_N(w,\pm,q)$
(resp. ${\cal N}^1_N(w,\pm,q)$), defined by the supplementary relations
\be b_i^2=1\quad\mbox{(resp. } (b_i+\sqrt{\pm1}w^{-1})(b_i+b_i^{-1})=0\mbox{ )},\label{ed1}\ee
where we recall that we used the convention $\sqrt{1}=1$ and $\sqrt{-1}=i$.
Finally, let us introduce
\be \label{rhrt}\hat r_{i,i+1}(\lambda)=e^\lambda\sigma_i-e^{-\lambda}\sigma_i^{-1} \qmbox{,} {\widetilde r}_{i,i+1}(\lambda)=w^{-1}\, e^{-\lambda}\rho_i-w e^{\lambda}\rho_i^{-1}\ \ee
and
\be \label{ktk} k_i(\lambda)=\begin{cases}b_i\ , \mbox{ for } {\cal N}^0_N(w,\pm,q)\\
e^\lambda w^{1\over 2} b_i+\eta e^{-\lambda}  w^{-{1 \over 2}} b_i^{-1}\ , \mbox{ for } {\cal N}^1_N(w,\pm,q)\, ,
\end{cases}\ee
where $\eta=\pm$ is a supplementary freedom.
These algebras allow us to obtain the following proposition solving essentially the Baxterisation problem in the case of the quantum twisted Yangian:
\begin{proposition}
The generators $\hat r_{i,i+1}(\lambda)$ satisfy the braided Yang-Baxter equation. We get also
\be
\hat r_{12}(\lambda_1-\lambda_2)\widetilde r_{23}(\lambda_1)\widetilde r_{12}(\lambda_2)
=\widetilde r_{23}(\lambda_2)\widetilde r_{12}(\lambda_2)\hat r_{23}(\lambda_1-\lambda_2)\, .
\ee
The generators ${\hat r}_{i,i+1}(\lambda)$, ${\widetilde r}_{i,i+1}(\lambda)$ and $k_i(\lambda)$ satisfy the braided reflection equation
\be
\hat r_{12}(\lambda_1-\lambda_2)k_1(\lambda_1)\widetilde r_{12}(\lambda_1+\lambda_2)
k_1(\lambda_2)=k_1(\lambda_2)\widetilde r_{12}(\lambda_1+\lambda_2)
k_1(\lambda_1)\hat r_{12}(\lambda_1-\lambda_2)\, .\label{retp}
\ee
We get also the following unitarity conditions
\be\label{unitrhrt}
\frac{\hat r_{12}(\lambda)\hat r_{12}(-\lambda)}{(qe^\lambda-q^{-1}e^{-\lambda})(qe^{-\lambda}-q^{-1}e^{\lambda})}=1~~~ {,}~~~~
\frac{{\widetilde r}_{12}(\lambda){\widetilde r}_{12}(-\lambda)}{(we^{\lambda}-w^{-1}e^{-\lambda})(we^{-\lambda}-w^{-1}e^{\lambda})}=1\, .
\ee
The unitarity condition for $k(\lambda)$ depends on the choice of the quotient: for ${\cal N}^0_N(w,\pm,q)$, we get $k_1(\lambda)k_1(-\lambda)=1$, for ${\cal N}^1_N(w,-,q)$,
\be
k_1(\lambda)k_1(-\lambda)=\eta e^{2\lambda}+\eta e^{-2\lambda}-w-w^{-1}\ ,
\ee
and, finally, for ${\cal N}^1_N(w,+,q)$,
\be
k_1(\lambda)k_1(-\lambda+i\frac{\pi}{2})=i\left(\eta e^{-2\lambda}-\eta e^{2\lambda}-w+w^{-1}\right) \, .
\ee
\end{proposition}
\textit{Proof.} This proposition is proved by direct computation expressing the generators ${\hat r}_{i,i+1}(\lambda)$, ${\widetilde r}_{i,i+1}(\lambda)$ and $k_i(\lambda)$ thanks to (\ref{rhrt})-(\ref{ktk}) and using the defining exchange relations (\ref{bd1})-(\ref{ed1}) of the algebras ${\cal N}_N(w,\pm,q)$, ${\cal N}^0_N(w,\pm,q)$ or ${\cal N}^1_N(w,\pm,q)$. For example, we get
\be
{\widetilde r}_{12}(\lambda){\widetilde r}_{12}(-\lambda)&=&
\left(w^{-1}\, e^{-\lambda}\rho_1-w e^{\lambda}\rho_1^{-1}\right)\left(w^{-1}\, e^{\lambda}\rho_1-w e^{-\lambda}\rho_1^{-1}\right)\non\\
&=&w^{-2}\rho_1^2+w^2\rho_1^{-2}-e^{-2\lambda}-e^{2\lambda}\non\\
&=&w^{-2}\rho_1(\rho_1-\rho_1^{-1})+w^{-2}-w^2\rho_1^{-1}(\rho_1-\rho_1^{-1})+w^2-e^{-2\lambda}-e^{2\lambda}\non\\
&=&w^{-2}+w^2-e^{-2\lambda}-e^{2\lambda}~,\qmbox{using relation (\ref{bd1})},\non
\ee
which proves the second relation of (\ref{unitrhrt}).
\finproof\\

The link between the generalized quantum Brauer algebra and the numerical previous solutions is given by the representation of ${\cal N}^0_N(w,\pm,q)$ or ${\cal N}^1_N(w,\pm,q)$. The map
\be
\label{repg1}
\sigma_i\mapsto \hat R_{i,i+1}\qmbox{,}\rho_i\mapsto \widetilde R_{i,i+1}\qmbox{and} b_i\mapsto V^t_i \mbox{  or  }i({\cal G}^-_q)_i\,V^t_i
\ee
is a representation of ${\cal N}^0_N(q^{n\over 2},(-1)^{n+1},q)$ (we recall that $q^{n \over 2}=e^{i\rho}$). It is proved by direct calculation. We then obtain the numerical matrices associated to the solutions (i) or (iv).
Similarly, we can prove that
\be
\label{repg2}
\sigma_i\mapsto \hat R_{i,i+1}\qmbox{,}\rho_i\mapsto \widetilde R_{i,i+1}\qmbox{and} b_i\mapsto {\widetilde {\cal K}}_i \mbox{ (given after (\ref{tks}))}
\ee
is a representation of ${\cal N}^1_N(q^{n\over 2},(-1)^{n+1},q)$.
In this case, we obtain the numerical solutions associated to solution (ii). Recall that we do not study the case (iii) since it provides a trivial finite algebra.

\subsection{Quantum Brauer duality}

We establish a result similar to the one of subsection 
\ref{secbrauer} but now for any generic solution $\cal K(\lambda)$ using the framework 
of the previous subsection.

We can define a finite algebra $\widetilde{\cal T}_f$ generated by $\widetilde \BB_0^\pm=\BB_0^\pm V_0^t$, which is obviously isomorphic to ${\cal T}_f$. These generators can be represented by
\be
\widetilde \BB_0^+\mapsto\widetilde{B}_0^+=B_0^+V_0^t=\hat R_{N0}\hat R_{N-1,N}\dots\hat R_{23}\hat R_{12}\ \widetilde{K}_1\
\widetilde{R}_{12}^{-1}\widetilde{R}_{23}^{-1}\dots\widetilde{R}_{N-1,N}^{-1}\widetilde{R}_{N0}^{-1}\\
\widetilde \BB_0^-\mapsto \widetilde{B}_0^-=B_0^-V_0^t=\hat R_{N0}^{-1}\hat R_{N-1,N}^{-1}\dots\hat R_{23}^{-1}\hat R_{12}^{-1}\ \widetilde{K}_1^{-1}\
\widetilde{R}_{12}\widetilde{R}_{23}\dots\widetilde{R}_{N-1,N}\widetilde{R}_{N0}
\ee
where $\widetilde{K}=V^t$ for (i), $\widetilde{K}=i{\cal G}^-_qV^t$ for (iv) and $\widetilde{K}$ is given just after (\ref{tks}) for (ii). Recall that we do not investigate the case (iii) which gives a trivial finite algebra. Let us remark that with this form of the algebra, we succeed in writing the generators $\widetilde B^\pm$ in terms of matrices appearing also in the representation
(see (\ref{repg1}) or (\ref{repg2})) of the algebra ${\cal N}_N(w,\pm,q)$.

The following proposition gives the analogue of the Brauer duality between the algebra generated by the elements of $\widetilde{B}_0^\pm$ and subalgebra of ${\cal N}_N(q^{n\over 2},\pm,q)$:
\begin{proposition}
\label{propduality}
Let us introduce $\displaystyle g_{1}=b_{1}^{-1}\ \frac{\rho_{1}-\rho_{1}^{-1}}{q-q^{-1}}\ b_{1}$.
The actions of the algebra generated by $\{\sigma_i|1\leq i \leq N-1\}$ and $g_{1}$ (see (\ref{repg1})-(\ref{repg2})) and of the algebra $\widetilde{\cal T}_f$ on the space $\left({\mathbb C}^n\right)^{\otimes N}$ commute with each other.
\end{proposition}
\textit{Proof.} The proof of this proposition is based in showing the following two relations, for $1\leq i \leq N-1$,
\be
\label{mproof}
\left[\,\hat R_{i,i+1}\, ,\, \widetilde B_0^\pm\, \right]=0\qmbox{and}
\left[\, {\widetilde{K}}_{1}^{-1}\left({\widetilde R}_{12}-{\widetilde R}_{12}^{-1}\right){\widetilde{K}}_1\, ,\, \widetilde B_0^\pm\, \right]=0
\ee
We shall use the fact that all the matrices involved in this computation satisfy the exchange relations of ${\cal N}_N(w,\pm,q)$. Let us give explicitly the proof for $[{\widetilde{K}}_1^{-1}\left({\widetilde R}_{12}-{\widetilde R}_{12}^{-1}\right){\widetilde{K}}_1,\widetilde B_0^+]=0$. Since ${\widetilde{K}}_1$ and ${\widetilde R}_{12}$ commute with matrices in space $0,3,4,\dots,N$, we can focus on proving
\be
\left[{\widetilde{K}}_1^{-1}\left({\widetilde R}_{12}-{\widetilde R}_{12}^{-1}\right){\widetilde{K}}_1,\hat R_{23}\hat R_{12}\ \widetilde{K}_1\
\widetilde{R}_{12}^{-1}\widetilde{R}_{23}^{-1}\right]=0\, .
\ee
Let us first compute
\be
{\widetilde{K}}_1^{-1}{\widetilde R}_{12}\ ({\widetilde{K}}_1 \hat R_{23})\ \hat R_{12} \widetilde{K}_1
\widetilde{R}_{12}^{-1}\widetilde{R}_{23}^{-1}
&=&{\widetilde{K}}_1^{-1}{\widetilde R}_{12}\hat R_{23}\ ({\widetilde{K}}_1\hat R_{12} \widetilde{K}_1
\widetilde{R}_{12}^{-1})\ \widetilde{R}_{23}^{-1}\non\\
&=&{\widetilde{K}}_1^{-1}\ ({\widetilde R}_{12}\hat R_{23}\widetilde{R}_{12}^{-1})\ {\widetilde{K}}_1\hat R_{12}
\widetilde{R}_{23}^{-1}\widetilde{K}_1\non\\
&=&{\widetilde R}_{23}^{-1}{\widetilde{K}}_1^{-1}\hat R_{12}{\widetilde{K}}_1\ (\widetilde{R}_{23}\hat R_{12}
\widetilde{R}_{23}^{-1})\ \widetilde{K}_1\non\\
&=&{\widetilde R}_{23}^{-1}\ ({\widetilde{K}}_1^{-1}\hat R_{12}{\widetilde{K}}_1 \widetilde{R}_{12}^{-1})\ \hat R_{23}
\widetilde{R}_{12} \widetilde{K}_1\non\\
&=&{\widetilde R}_{23}^{-1}\widetilde{R}_{12}^{-1}{\widetilde{K}}_1\hat R_{12}\hat R_{23}\ \
{\widetilde{K}}_1^{-1}\widetilde{R}_{12}\widetilde{K}_1
\ee
In the latter relations, the parenthesis indicate which part of the product is modified using different relations (\ref{bd1})-(\ref{ed2}). We obtained similar result for ${\widetilde{K}}_1^{-1}{\widetilde R}_{12}^{-1}{\widetilde{K}}_1$. Then, we have proved that
\be
\label{presque}
{\widetilde{K}}_1^{-1}\left({\widetilde R}_{12}-{\widetilde R}_{12}^{-1}\right){\widetilde{K}}_1{\widetilde B}_0^+={\widetilde C}_0^+
{\widetilde{K}}_1^{-1}\left({\widetilde R}_{12}-{\widetilde R}_{12}^{-1}\right){\widetilde{K}}_1,
\ee
where $\widetilde C_0^+=\hat R_{N0}\hat R_{N-1,N}\dots\hat R_{34}\widetilde{R}_{23}^{-1}\ \widetilde{R}_{12}^{-1}\widetilde{K}_1\
\hat R_{12}\hat R_{23}\widetilde{R}_{34}^{-1}\dots\widetilde{R}_{N-1,N}^{-1}\widetilde{R}_{N0}^{-1}$. Multiplying on the left by ${\widetilde{K}}_1^{-1}{\widetilde R}_{12}{\widetilde{K}}_1$ the previous relation and knowing that it satisfies relation (\ref{bd1}), we finish the proof.
\finproof\\
Within this framework, the proof of this proposition requires only
the exchange relations of the algebra ${\cal N}(w,\pm,q)$ whereas,
in the previous case (see subsection \ref{secbrauer}), the proof
of the analogous proposition is based on direct computation.

\noindent{\bf Acknowledgments:} One of the authors (N.C.) wishes to thank the Physics Department of the University of Bologna and the INFN section of Bologna, where this work was initiated. This work was supported by INFN and the European Network `EUCLID'; `Integrable models and applications: from strings to condensed matter',
contract number HPRN--CT--2002--00325.

\appendix

\section{Appendix}\label{appexp}

In this appendix, we give the explicit expression of the commutation relations between
the finite algebra and the quantum twisted Yangian for ${\cal U}_q(gl_3)$ and for ${\cal K}(\lambda)={\mathbb I}$.
Let us express ${\mathbb B}$ as a matrix \be {\mathbb B}(\lambda) =
\left(
\begin{array}{ccc}
 {\cal A}_{1}  &{\cal D}_{1}  & {\cal D}  \\
 {\cal C}_{1}  &{\cal A}_{2}  & {\cal D}_{2} \\
 {\cal C}      &{\cal C}_{2}  & {\cal A}_{3}  \\
\end{array} \right). \label{tr2} \ee
Set also $c = q-q^{-1}$ and $[X,\ Y]_{q} = q X Y - q^{-1} Y
X$. We recall that the diagonal charges are trivial
$B_{ii}^{+} \propto {\mathbb I}$ so we simply deal with the
rest of them.  Then from (\ref{exc1}) and bearing in mind (\ref{tr2}) various sets of
equations are inferred:
\be \Big [B_{12}^{+}, {\cal A}_1 \Big ]_{q} &=& q c
{\cal C}_{1} +c {\cal D}_{1}, ~~~~~\Big [B_{12}^{+},
{\cal A}_2 \Big ]_{q^{-1}} = -q c {\cal C}_{1} - c\ {\cal D}_{1},
~~~~~[B_{12}^{+}, {\cal A}_3 \Big ]=0, \non\\
q \Big [B_{12}^+,\ {\cal D}_{1} \Big ] &=& \Big [B_{12}^+,\ {\cal C}_{1} \Big ] =q c ({\cal A}_{2} -{\cal A}_{1}),
~~~~ \Big [B_{12}^{+},\ {\cal D}_{2}\Big ]_{q^{-{1\over
2}}}  = -q^{{1\over 2}}c {\cal D} \non\\ \Big [B_{12}^{+},\
{\cal C}_{2} \Big ]_{q^{-{1\over 2}}} &=&-q^{{1\over 2}} c {\cal
C}_{1}, ~~~~~ \Big [B_{11}^{+},\ {\cal D}\Big
]_{q^{{1\over 2}}} = q^{1\over 2} c {\cal D}_{2}, ~~~~\Big [B_{12}^{+},\ {\cal C}\Big ]_{q^{{1 \over 2}}} = q^{{1\over 2}}c
{\cal C}_{2} \ee \be \Big [B_{23}^{+},\ {\cal A}_{1} \Big
] &=& 0, ~~~~\Big [B_{23}^{+},\ {\cal A}_{2} \Big ]_{q} =
q\ c\ {\cal C}_{2} +c{\cal D}_{2}, ~~~~[B_{23}^{+},\ {\cal
A}_{3} \Big ]_{q^{-1}} = -q c {\cal C}_{2} -c\ {\cal D}_{2} \non\\
q \Big [B_{22}^+,\ {\cal D}_{2} \Big ] &=& \Big [B_{23}^+,\ {\cal C}_{2} \Big ]=  q c  ({\cal A}_{3}-{\cal
A}_{2}), ~~~~\Big [B_{23}^+,\ {\cal D}_1\Big ]_{q^{{1\over
2}}} =q^{{1\over 2}} c {\cal D}\non\\ \Big [B_{23}^+,\
{\cal C}_{1}\Big ]_{q^{{1\over 2}}} &=& q^{{1\over 2}} c {\cal C},
~~~~\Big [B_{23}^+,\  {\cal D} \Big ]_{q^{-{1\over 2}}}
=-q^{{1\over 2}} c {\cal D}_{1},  ~~~~\Big[B_{23}^+,\
{\cal C}\Big ]_{q^{-{1\over 2}}} =-q^{{1\over 2}} c {\cal C}_{1}
\ee \be \Big [B_{13}^{+},\ {\cal A}_{1} \Big ]_{q} &=& q c
{\cal C} +c {\cal D}, ~~~~\Big [B_{13}^{+},\ {\cal A}_{3}
\Big ]_{q^{-1}} = -q c {\cal C} -c\ {\cal D}, \non\\ q \Big [B_{13}^{+},\ {\cal D} \Big ] &=& \Big [B_{13}^{+},\ {\cal C} \Big ]=  q c  ({\cal A}_{3}-{\cal A}_{1}), ~~~~ \Big [B_{13}^{+},\ {\cal D}_1 \Big ]_{q^{{1\over 2}}}  =
c (q^{{1\over 2}}{\cal C}_{2} + q^{-{1\over 2}}{\cal D} B_{12}^+ - q^{-{1\over 2}}{\cal A}_{1}B_{23}^{+}) \non\\ \Big [B_{13}^+,\  {\cal C}_{1}\Big
]_{q^{{1\over 2}}}  &=& q^{{1\over 2}}c ({\cal D}_{2} +B_{12}^+ {\cal C} -B_{23}^+{\cal A}_{1}), ~~~~~\Big[ B_{13}^+,\ {\cal D}_{2}\Big ]_{q^{-{1\over 2}}}= c(-q^{{1\over 2}}
{\cal C}_{1} +q^{-{1\over 2}} B_{12}^+ {\cal A}_{3}
-q^{-{1\over 2}} B_{23}^+ {\cal D}), \non\\ \Big[B_{13}^+,\ {\cal C}_{2}\Big ]_{q^{-{1\over 2}}} &=& q^{{1\over
2}}c({\cal A}_{3} B_{12} - {\cal C}B_{23} - {\cal
D}_{1}), ~~~~\Big [B_{13}^{+},\ {\cal A}_{2} \Big ] = c(- B_{23}^+ {\cal
D}_{1} -{\cal C}_{1} B_{23}^{+} +{\cal D}_{2} B_{12}^{+} +B^+_{12}{\cal C}_{2}). \non \ee

\end{document}